\documentclass[fleqn,10pt]{wlscirep}
\title{Wealth and Identity: The dynamics of dual segregation}

\author[1,*]{Anand Sahasranaman}
\author[1]{Henrik Jeldtoft Jensen}
\affil[1]{Department of Mathematics and Centre for Complexity Science, Imperial College London, London, United Kingdom}

\affil[*]{a.sahasranaman15@imperial.ac.uk}

\begin{abstract}
We extend our model of wealth segregation to incorporate migration and study the tendencies towards `dual' segregation -  segregation due to identity (migrants vs. residents) and segregation due to wealth. We find a sharp, non-linear transformation between segregated and mixed-wealth states as neighborhood wealth thresholds become less stringent, allowing agents to move into neighborhoods they cannot afford.  The number of such moves required for the onset of this transformation varies inversely with the likelihood of agents willing to move into less wealthy neighborhoods. We also find that this sharp transformation from segregated to mixed wealth states is simultaneously accompanied by a corresponding non-linear transformation from a less identity-segregated to a highly identity-segregated state. We argue that the decrease in wealth segregation does not merely accompany, but in fact drives the increase in identity-based segregation. This implies that lowering wealth segregation necessarily exacerbates identity segregation and that therefore, this trade-off could to be an important consideration in designing urban policy. Additionally, we find that the time taken for the tolerance levels of residents to be breached is significantly lesser under rapid migration. This rapidity of breach of tolerance could potentially underpin the intensity of resident responses to sharp bursts of migration.
\end{abstract}
\begin{document}

\flushbottom
\maketitle

\thispagestyle{empty}

\section*{Introduction}

Cities around the world are dealing with the phenomenon of increased migration. The UN projects that 66\% of the global population will live in cities by 2050, up from 54\% in 2014, and that 90\% of this increase is expected to be focused in Asia and Africa~\cite{bib0a}. In addition to this expected migration, we also see continued forced displacement occurring due to climate change on the one hand, and wars and conflicts on the other. Myers~\cite{bib0b}, for instance, estimates that there could be as many as 200 million people displaced worldwide on account of global warming by 2050. Therefore, one of the central questions confronting cities as they plan for the future relates to the appropriate strategies to adopt in dealing with increasing influx of migrants while still maintaining social and economic stability.

We had previously investigated the dynamics underlying the emergence and reversal of wealth based segregation in cities~\cite{bib0b2} as well as the longer term evolution of economic status of urban neighborhoods over time~\cite{bib0b3}. The basis of this work is the Schelling model~\cite{bib1}, the classic framework to study segregation. It explains the emergence of racial segregation on the basis of interactions of individual agent preferences for `like' neighbors, with even small preferences for like neighbors at an individual level leading to emergence of segregated patterns at a collective level. The striking finding of the Schelling model is that even if all agents have a high tolerance for agents unlike themselves (up to two-thirds of their neighbors), large scale segregation ensues. The Schelling model is robust across parameter specifications such as neighborhood definition, heterogeneity of agent preferences, and agent choice functions~\cite{bib2b}\textsuperscript{,}~\cite{bib2c}\textsuperscript{,}~\cite{bib2d}\textsuperscript{,}~\cite{bib2e}\textsuperscript{,}~\cite{bib2f}. We had modeled wealth based segregation~\cite{bib0b2}, modifying the Schelling model~\cite{bib1} by replacing agent specific tolerance levels with neighborhood level wealth threshold levels (based on the wealth profile of extant agents in the neighborhood) that agents were required to satisfy in order to be able to move into these neighborhoods. We found a sharp transition from segregated to mixed wealth equilibrium when only a small fraction of agents were allowed to move into neighborhoods in contravention of the neighborhood wealth condition. This effect appears to be consistent with the findings of Benenson, Hatna, and Or~\cite{bib2g}, whose empirical work on Israeli cities suggested that the introduction of a small number of highly tolerant agents could result in significantly mixed residential patterns. It is important to note that there is a substantial difference in interpretation of the observed phenomenon, with our work positing the mixed equilibrium as an outcome of even a limited contravention of wealth threshold conditions presented by the city, while Benenson, Hatna, and Or present a classical Schelling interpretation of the phenomenon based on agent tolerances. Sethi and Somanathan~\cite{bib0k} develop a theoretical framework where residential choice of agents is determined both by the affluence and racial composition of neighborhoods. They find that extreme racial segregation can be a stable configuration if racial income disparities are either large or small and that when racial income disparities are small there are in fact multiple stable equilibria representing dramatically different levels of racial segregation.

While there have been diverse explorations of the Schelling model in the literature such as through agent based simulations in sociology~\cite{bib2h} and economics~\cite{bib2i}, or in the study of the phase transition in the model using tools of statistical mechanics~\cite{bib2j}\textsuperscript{,}~\cite{bib2k}, the question of migration into a Schelling framework appears to be a significant aspect that has been neglected. Urselmans' recent work~\cite{bib0b4} appears to be the first instance of exploring the effect of migration on identity based segregation. In Urselmans' model, the migrating population belongs to a different identity (ethnic/racial) group from the extant population of the city, and she finds that overall agent happiness (in terms of fraction of co-ethnics in the neighborhood) converges, irrespective of the size and rate of migration. However, population density is found to amplify segregation and impact agent happiness. 

There has been a significant body of sociological literature that attempts to understand where migrants go when they enter the city. For instance, the theory of spatial assimilation~\cite{bib0c} posits that immigrants first settle in homogeneous ethnic enclaves both due to the comfort of being amongst co-ethnics and because of the inability of many newcomers to afford living in neighborhoods of the majority. The theory further contends that as their socioeconomic condition improves, migrants seek to improve their spatial location and move into wealthier neighborhoods of the dominant ethnic majority. Analysis of census data for London reveals distinct neighborhoods for immigrants of different ethnicities~\cite{bib0b4}, and detailed ethnographies, such as by Hall~\cite{bib0e}, further confirm that ethnic diversity is concentrated by area, with migration a primary contributor. Novotny and Hasman~\cite{bib0f} find that spatial relatedness between incoming immigrants and existing communities is a useful measure in analyzing the dynamics of regional population change.

Yet another significant aspect of the population migrating into cities is their wealth, which dictates their ability to make the residential choices they desire. Cities are sinks of migration because they are seen by migrants as magnets of economic growth and pathways to a better life by enabling them to earn and save more. Glaeser~\cite{bib0fa} points to the strong correlation between urbanization and economic development across nations and also to the increased productivity within nations enabled by the agglomeration economies facilitated by urbanization. There is evidence to suggest that, on average, incoming rural migrants have lower wealths than urban dwellers. For instance, in India where 200 million people are expected to move from villages to cities between 2011 and 2030~\cite{bib0g}, Zacharias and Vakulabharanam~\cite{bib0h} find that mean wealth in urban India is 1.74 times that of rural India. They also find that wealth is stratified by caste, with historically disadvantaged castes - the Scheduled Castes and Scheduled Tribes - being economically worse off than all other caste groups in both rural and urban contexts. This group also forms a significant proportion, close to 45\%, of the current rural population and consequently the bulk of expected urban migrants.Therefore, it is likely that a new migrant into a city will be economically and socially dominated by extant city dwellers. This is also evidenced in cases where the migration is forced by crisis in a host nation, and as a consequence there is a sudden, sharp influx of refugees and asylum seekers into destination cities. For example, on account of the crisis in Syria, Germany received over 475,000 asylum applications in 2015, but even more dramatically, saw the arrival of more than 1 million refugees into its cities that year~\cite{bib0i}. A socioeconomic survey of Syrian refugees~\cite{bib0j} finds that most refugees worked in low-paying and insecure jobs, evidence of their precarious economic condition and low wealth status.  

The entry of migrants has also been found to impact the attitudes of extant city residents to the new arrivals. For instance, Wike, Stokes and Simmons~\cite{bib0ka} find that post the recent surge of Syrian refugees in Europe, over half the population in eight out of ten European countries surveyed believe that incoming refugees increase the likelihood of terrorism in their country and over half the population in six of the countries see refugees as a burden on jobs and social benefits. The rise of these attitudes, they argue, can be linked to the growing anti-immigrant rhetoric amongst political parties in Europe. Milanovic further substantiates this phenomenon by providing evidence of the significant and rapid increase in vote share of populist, nativist political parties in Europe between 2000 and 2017~\cite{bib0kb}. It is therefore plausible that the rapidity of migration could indeed impose a significant influence on the intensity of resident attitudes to migrants - a question for deeper exploration.

As outlined above, migration and population increase are significant drivers of city evolution, especially patterns of urban settlement, and in view of this, we now propose to expand our work on wealth segregation to incorporate migration. Our exploration seeks to extend the nascent work on migration and segregation in the Schelling framework, by investigating the dynamics of agent migration and movement based on the dual drivers of wealth and identity. Specifically, we are interested in studying the interplay between the dynamics of agent migration in enabling new populations to move into the city and agent wealth in mediating movement within the city, and the consequent relationship between the tendencies towards `dual' segregation -  segregation due to identity (migrants vs. extant residents) and segregation due to wealth. We also study the effect of varying the nature of migration (gradual vs. rapid) on the tolerance levels of existing residents in the city.

\section*{Model Definition and Specifications}
We consider a `city' with $M$ neighborhoods, and each neighborhood $i$ ($i$ $\in$ $1$,...,$M$), is populated by $P(i)$ agents (households). Initially, agents are distributed equally across each of the M neighborhoods, such that $P(i) = P(j)$ for $1 \le i,j \le M$. The initial set of agents are defined to  be the original inhabitants of the city - termed `original agents'. Each original agent has two attributes - wealth and migrant tolerance level. We sample agent wealths from a normal distribution - $N(10,1)$. The total wealth of each neighborhood $i$ is the sum of individual agent wealths and is denoted by $W\textsubscript{tot}(i)$. As with the Tolerance level in the Schelling model, we define the migrant tolerance level ($\tau$) as the fraction of migrant population in a neighborhood that an original agent is happy to tolerate. Sethi and Somanathan~\cite{bib0k} state that studies on preferences for neighbors have historically shown that people prefer some degree of integration with a bias for own-type neighbors. They point, for instance, to a survey of 8,500 households on urban inequality from the 1990s~\cite{bib0ja} which finds that, on average, African Americans preferred neighborhoods with 40\% like neighbors, while white Americans preferred neighborhoods with 52\% white neighbors. We set $\tau = 0.5$ for all original agents in our simulations. We also simulate the dynamics for a broader set of $\tau$ values in the Supplementary Information appendix (SI.1).

The dynamics of the model consist of two distinct aspects - migration into the city and movement within the city. 

At each iteration of the model, a certain number of agents, defined as a fraction of extant city population, attempt to enter the city. Given this fraction, $r\textsubscript{mig}$, and the total population of the city at the end of the previous iteration ($P\textsubscript{t-1}$), the number of agents attempting to migrate into the city at time step $t$, $Mig\textsubscript{t}$, is determined as (Eq~\ref{eq:migpop}): 

\begin{eqnarray}
\label{eq:migpop}
	\mathrm{Mig_{t}} = P_{t-1} (1 + r_{mig})
\end{eqnarray}

\noindent
For our simulations of the model, we use two scenarios - first, a gradual population increase over each iteration of the model and second, a rapid population increase over a much shorter set of iterations. We do this in order to mimic both orderly, predictable growth over time as well as bursty growth over short time spans that could, for instance, be caused by crises such as climate change and conflict. For the gradual growth case, $r\textsubscript{mig}$ = 0.005 over the 500 iterations of the model, and for the rapid growth case, $r\textsubscript{mig}$ = 0.025 over the first 100 iterations and $r\textsubscript{mig}$ = 0.0 over the remaining 400 iterations of the model. 

While $Mig\textsubscript{t}$ agents attempt to enter the city at $t$, the actual number that are able to enter the city is contingent upon the wealth of the agent and the fraction of original agents to migrant agents in city neighborhoods. Given our discussion earlier that the bulk of migrants will probably be less wealthier, on average, than agents resident in the city, we sample migrant agent wealths from a normal distribution with lower mean wealth than the wealth distribution for original agents - $N(7,1)$. However, for the sake of comparison, we also sampled migrant agent wealths from the same distribution as original agent wealth and find that the outcomes are similar to those presented here. We present the details of this analysis in the Supplementary Information appendix (SI.2).

For each migrating agent (out of $Mig\textsubscript{t}$ agents) whose wealth is greater than or equal to the median wealth of a randomly chosen city neighborhood $i$, entry into that neighborhood is guaranteed - ie. probability of agent entry, $p\textsubscript{entry}$ = $1$. In case of the remaining agents for whom the wealth threshold condition is not satisfied, entry into the city is stochastically determined, based on the ratio of population of original agents ($P\textsubscript{i}\textsuperscript{O}$) to total population ($P\textsubscript{i}\textsuperscript{Tot}$) in the randomly chosen cell $i$ (Eq~\ref{eq:p_entry}). 

\begin{eqnarray}
\label{eq:p_entry}
	\mathrm{p_{entry}} = \exp (-\beta_{in} \frac{P_i^O}{P_i^{Tot}})
\end{eqnarray}

\noindent
This is designed to mimic the empirically observed phenomenon of identity based spatial clustering of new migrants we had discussed in the previous section. Therefore, the probability of an agent being able to enter a neighborhood increases with the increase in proportion of migrant population to original population, even when in contravention of the wealth threshold condition. In Eq~\ref{eq:p_entry}, $\beta\textsubscript{in}$ is simply a calibrating factor that determines $p\textsubscript{entry}$. In the limit $\beta\textsubscript{in}\rightarrow\infty$, the probability of migrant entry becomes purely dependent on satisfaction of the wealth threshold condition, while in the limit $\beta\textsubscript{in}\rightarrow0$, all potential $Mig\textsubscript{t}$ migrants enter the city. For the simulations of the model, we choose $\beta\textsubscript{in}$ = 1. To compare, we also conduct the analysis for $\beta\textsubscript{in}$ = 10 in the Supplementary Information appendix (SI.3), and find that the results were analogous to those presented here. Once all eligible migrants have entered the city for a given iteration (time step $t$), the population of the city ($P\textsubscript{t}$) is the sum of the population of agents at time step $t-1$, $P\textsubscript{t-1}$, and the new migrant population that has successfully entered at time step $t$.

After the migration dynamics are completed for an iteration, we shift focus to agent movement for that iteration (time step $t$). Agent movement occurs as a two step process - first, the decision of an agent in choosing to move from or stay in its current location and second, if it does choose to move, the ability of the agent to accomplish the move given its wealth. The agent decision  step involves two components, namely the assessment of the wealth of its current and potential locations as well as the composition of agent identities in these neighborhoods.

Agents decide on wanting to move or stay, firstly, by taking stock of the relative wealthiness of their neighborhood to other neighborhoods in the city. The choice of movement based on such a neighborhood wealth comparison is supported by empirical evidence on the link between the economic status of neighborhoods and a variety of socioeconomic outcomes. For instance, Chetty and Hendren~\cite{bib0l} find that neighborhoods affect inter-generational mobility and that outcomes of children moving into a better neighborhood improve linearly with the time spent in that area. They also find that living in a higher income neighborhood could have significant positive impact on future incomes of children from lower income households. Similarly, Chyn~\cite{bib0m} cites significant positive impact on employment and income levels of children moving out of disadvantaged neighborhoods. Our previous work~\cite{bib0b3} also finds that the use by agents of a neighborhood comparison heuristic to determine their movement decision is critical to generating long term patterns of empirically observed change in the economic status of neighborhoods in cities. 

At time step $t$, we randomly choose $P\textsubscript{t}$ agents sequentially, and each of them then makes the choice of moving within the city. The choice of $P\textsubscript{t}$ agents at  $t$ ensures that, on average, every agent has the opportunity to move at each time step. For each such randomly chosen agent, $A$, in neighborhood $i$, a random receiving cell $j$ is chosen and the neighborhood comparison condition is applied - that is, the median wealths of these neighborhoods $i$ and $j$ are compared. If the median wealth of the receiving cell $j$ ($W\textsubscript{Med}\textsuperscript{j}$) is greater than that of the agent's current location $i$ ($W\textsubscript{Med}\textsuperscript{i}$), then the agent chooses to attempt a move from its current location, ie. $p\textsubscript{init}\textsubscript{\_}\textsubscript{choice} = 1$. If the median wealth of $i$ is greater than that of $j$, the agent would, in general, prefer to stay in its current location, but we model an element of stochasticity in this choice - to capture other aspects that might influence such a decision (Eq~\ref{eq:probchoice}). For instance, in case of an original agent, this stochasticity could capture the conflict inherent in weighing the relative merits of neighborhood wealth profiles and neighborhood identity profiles in choosing to stay or attempting to move.

\begin{eqnarray}
\label{eq:probchoice}
	\mathrm{p_{init\_choice}} = \exp (-\beta_{choice}(W_{Med}^{i} - W_{Med}^{j} ))
\end{eqnarray}

\noindent
$\beta\textsubscript{choice}$ is the calibrating factor that determines $p\textsubscript{init}\textsubscript{\_}\textsubscript{choice}$ . In the limit $\beta\textsubscript{choice}\rightarrow\infty$, choice is completely deterministic as the only agents that choose to attempt a move are those for whom $W\textsubscript{Med}\textsuperscript{j}\geq W\textsubscript{Med}\textsuperscript{i}$. In the limit $\beta\textsubscript{choice}\rightarrow0$, choice is again completely deterministic as all agents choose to attempt a move. For our model, we choose $\beta\textsubscript{choice}$ = 10, to allow for some movement in  breach of the neighborhood comparison condition. For comparison, we also run the dynamics for $\beta\textsubscript{choice} = 2$ and $\infty$, and analyze the outcomes in the Supplementary Information appendix (SI.4).

The second step in the agent's decision making process requires an assessment of the identity of agents in its current and receiving cells. Again, here we draw a distinction between the original agents (who occupied the cell at $t$ = 0) and migrant agents, who have been immigrating into the city after $t$ = 0. We now incorporate the identity based effects, as in the classic Schelling model, for the original agents who are defined to have a migrant tolerance level of $\tau$. We do not model a contrasting tolerance level for the migrating agents in view of our earlier discussion on the lower wealth status, on average, of immigrants and their desire to move into the better neighborhoods of the dominant community over time. 

Given that $p\textsubscript{init}\textsubscript{\_}\textsubscript{choice}$ is realized for a randomly chosen original agent, the next comparison is between the fraction of migrants in the two cells. If the fraction of migrants in the receiving cell is lesser than either $\tau$ or the fraction of migrants in the agent's current cell, then the agent chooses to attempt a move with probability 1, ie. $p\textsubscript{final}\textsubscript{\_}\textsubscript{choice} = 1$. If, however, the fraction of migrants in the receiving cell is greater than both the fraction of migrants in its current location and $\tau$, $p\textsubscript{final}\textsubscript{\_}\textsubscript{choice} = 0$. For all migrant agents seeking to move based on the realization of $p\textsubscript{init}\textsubscript{\_}\textsubscript{choice}$, $p\textsubscript{final}\textsubscript{\_}\textsubscript{choice}$ = 1.

Finally, if $p\textsubscript{final}\textsubscript{\_}\textsubscript{choice}$ is realized and the agent has chosen to attempt a move out of its current location, then the actual occurrence of the move is determined by the agent's wealth. If the agent's wealth ($W\textsubscript{A}$) is greater than or equal to the median wealth of the receiving cell ($W\textsubscript{j}\textsuperscript{Med}$), then the agent moves to the receiving neighborhood with probability $p\textsubscript{move} = 1$. If the wealth threshold condition is not satisfied, the agent move becomes stochastic, with probability $p\textsubscript{move}$ (Eq~\ref{eq:probmove}):

\begin{eqnarray}
\label{eq:probmove}
	\mathrm{p_{move}} = \exp (-\beta_{move}(W_{j}^{Med} - W_{A}))
\end{eqnarray}

\noindent
$\beta\textsubscript{move}$ is the calibrating factor that determines $p\textsubscript{move}$ . In the limit $\beta\textsubscript{move}\rightarrow\infty$, movement is completely deterministic as the only moves that occur are those that satisfy the threshold wealth condition. As we explored earlier~\cite{bib0b2}, decreasing $\beta\textsubscript{move}$, thereby enabling moves in contravention of the threshold wealth condition, yields a sharp transformation from a highly segregated to a mixed wealth configuration even at a very small fraction of such moves (termed as disallowed-realized moves). We now seek to study the impact on identity based segregation of a progressively decreasing $\beta\textsubscript{move}$, while also verifying whether the sharp transformation from segregated to mixed wealth states still occurs as previously observed. For this analysis, $\beta\textsubscript{move}$ takes the following values: 1000, 100, 10, 5, 1, 0.1, 0.01, and 0.001.

In summary, the update algorithm for each iteration consists of two components - agent migration and agent movement. Agents migrate into the city on the basis of their wealth and if their wealth is insufficient, then on the fraction of original agent populations in the randomly chosen neighborhoods they seek to move into - captured in the realization of $p\textsubscript{entry}$. Following this, agents are chosen at random and make decisions on movement. By design, the number of random agents chosen in each iteration is equal to the population of the city in that iteration so that all extant agents, on average, make movement decisions in every time step. Every randomly chosen agent at each iteration chooses to move from or stay in its neighborhood based on the comparative median wealths of current and prospective neighborhoods, captured in the realization of $p\textsubscript{init}\textsubscript{\_}\textsubscript{choice}$. Original agents also incorporate the identity make up of both neighborhoods and their migrant tolerance levels in their choice to move. Migrant agents' choice of movement is based solely on the neighborhood wealth comparison. Once the final choice is made to attempt a move, based on realization of $p\textsubscript{final}\textsubscript{\_}\textsubscript{choice}$, the actual movement in case of both migrant and original agents is based on the difference between their agent wealths and the median wealth of the receiving cell, which is captured in the realization of $p\textsubscript{move}$.

The update process in the model is sequential, which means that all model variables are updated at the end of each agent action. We model a city with $M$ = 50 neighborhoods, but for the sake of comparison provide the analysis for $M$ = 25 in the Supplementary Information appendix (SI.5) where we find that the outcomes are similar to those described here. The dynamics are run over 500 iterations, which means that each original agent, on average, is sampled 500 times, and all migrant agents, on average, are sampled as many times as the number of time steps (iterations) they spend in the city. We define 10 iterations or time steps as 1 time sweep; therefore the 500 iterations correspond to 50 time sweeps of the dynamics. Finally, for each combination (as detailed in Table~\ref{table1}) of parameter values of rate of migration ($r\textsubscript{mig}$) and $\beta\textsubscript{move}$, we run an ensemble of 12 runs of the entire model.

\begin{table}[!ht]
\centering
\captionsetup{justification=centering}
\caption{\bf Model Parameters.}
\begin{tabular}{|l|l|}
\hline
Number of Neighborhoods ($M$) & 50\\ \hline
Number of Original Agents & 2,500\\ \hline
Migrant Tolerance Level ($\tau$) & 0.5 \\ \hline
Migration Rate ($r\textsubscript{mig}$) & Gradual: 0.005 over 500 iterations; \\& Rapid: 0.025 over 100 iterations and 0.0 over remaining 400 iterations\\ \hline
Agent Wealth Distribution & Original Agents: $N(10,1)$; \\& Migrant Agents: $N(7,1)$\\ \hline
Number of Iterations & 500 \\ \hline
Number of Time Sweeps & 50 \\ \hline
$\beta\textsubscript{in}$ & 1 \\ \hline
$\beta\textsubscript{choice}$ & 10 \\ \hline
$\beta\textsubscript{move}$ & 0.001, 0.01, 0.1, 1, 5, 10, 100, 1000 \\ \hline
\end{tabular}
\label{table1}
\end{table}

Given our objective of understanding the potential emergence of `dual' segregations - wealth and identity based - we seek to measure the following outcomes at the end of the 50 time sweeps: Average Fraction of Rich Neighbors ($F\textsubscript{R}$) and Average Size of Original Agent Neighborhoods ($S\textsubscript{O}$). In order to measure the impact of the nature of migration on original agent tolerance, we define two outcomes: Average Time to Tolerance Breach ($T\textsubscript{$\tau$}$) and Fraction of Original Agents with Tolerance Breach ($F\textsubscript{$\tau$}$).

Average Fraction of Rich Neighbors ($F\textsubscript{R}$) is simply a measure of the average fraction of rich neighbors of a rich agent. For the purposes of our analysis, a rich agent is defined as an agent with wealth in the top 15\% of the population. Let $N\textsubscript{R}$ be the total number of rich agents in the city. For each rich agent $a\textsubscript{R}$, we compute the number of its rich neighbors ($V\textsubscript{R}(a\textsubscript{R})$) as a fraction of its total neighbors ($V(a\textsubscript{R})$) and average this quantity across all $N\textsubscript{R}$ rich agents (Eq~\ref{eq:Fr}). 

\begin{eqnarray}
\label{eq:Fr}
	\mathrm{F_R} = \frac{1}{N_R} \sum_{a_{R}=1}^{N_R} \frac {V_R(a_R)} {V(a_R)}
\end{eqnarray}

Average Size of Original Agent Neighborhoods ($S\textsubscript{O}$) is a measure of the average number of original agents that are neighbors of original agents. Let $N\textsubscript{O}$ be the total number of original agents in the city. For each original agent $a\textsubscript{O}$, we compute the number of its neighbors that are also original agents ($V\textsubscript{O}(a\textsubscript{O})$) and average this across all $N\textsubscript{O}$ original agents (Eq~\ref{eq:So}):  

\begin{eqnarray}
\label{eq:So}
	\mathrm{S_{O}} = \frac {1}{N_O} \sum_{a_{O}=1}^{N_O} {V_O(a_O)}
\end{eqnarray}

The Average Time to Tolerance Breach ($T\textsubscript{$\tau$}$) is the number of time sweeps, on average, taken for an original agent to have its migrant tolerance limit ($\tau$) breached for the first time. For each original agent $a\textsubscript{O}$, we compute the number of time sweeps taken for the agent to reach its migrant tolerance limit ($T\textsubscript{$\tau$}(a\textsubscript{O}$)) and average this across all $N\textsubscript{O}(\tau)$ original agents whose migrant tolerance limits have been breached by the end of the simulations (Eq~\ref{eq:Ttau}).

\begin{eqnarray}
\label{eq:Ttau}
	\mathrm{T_{\tau}} = \frac {1}{N_{O}(\tau)} \sum_{a_{O}=1}^{N_{O}(\tau)} {T_{\tau}(a_O)}
\end{eqnarray}

Fraction of Original Agents with Tolerance Breach ($F\textsubscript{$\tau$}$) is the number of original agents whose migrant tolerance levels have been breached by the end of the simulations as a fraction of total number of original agents (Eq~\ref{eq:Ftau}). 

\begin{eqnarray}
\label{eq:Ftau}
	\mathrm{F_{\tau}} = \frac {N_{O}(\tau)} {N_{O}}
\end{eqnarray}

\section*{Results}
\begin{figure}[ht]
\centering
\includegraphics[width=\linewidth]{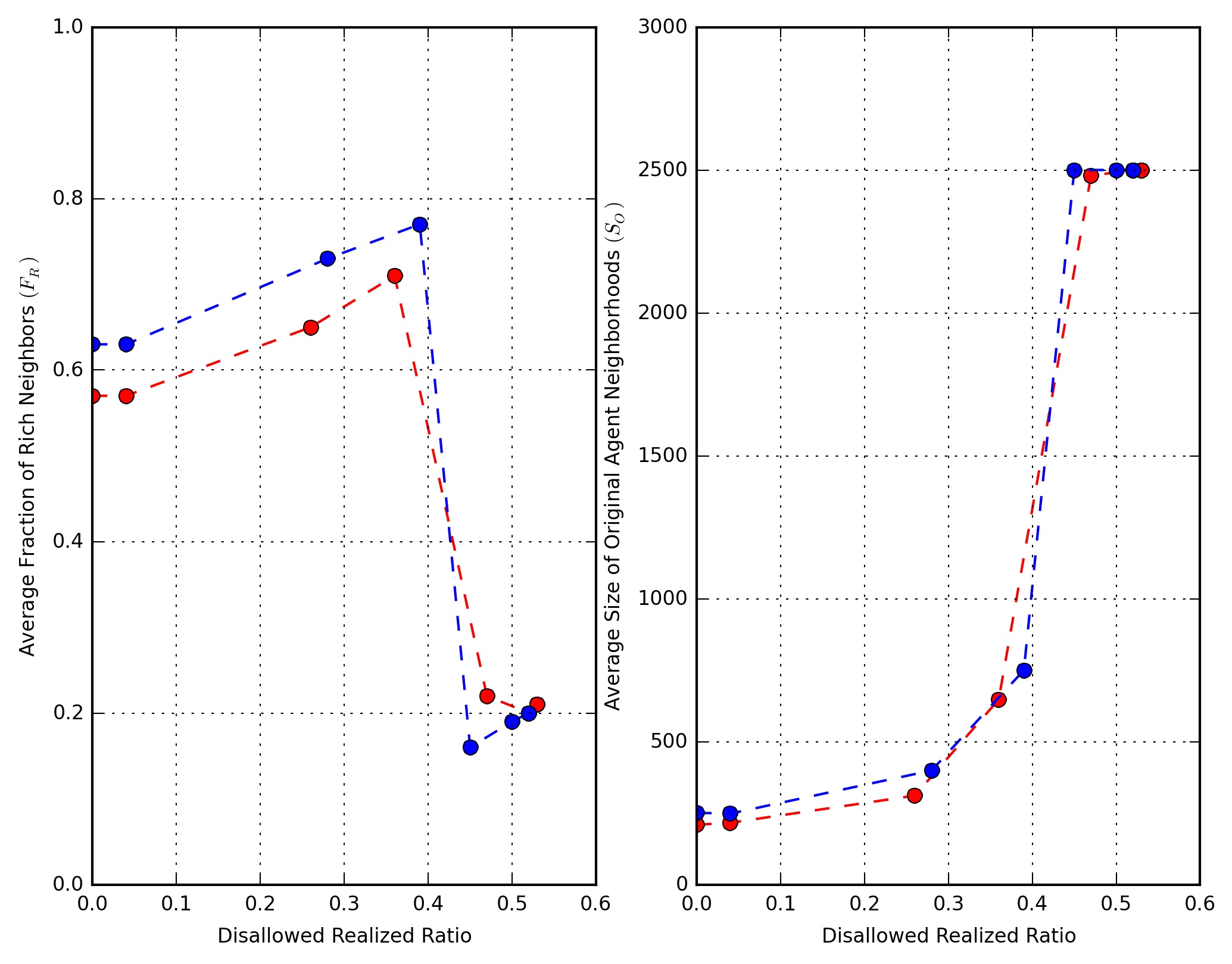}
\caption{Emergence of `dual' segregations. LEFT: Average Fraction of Rich Neighbors ($F\textsubscript{R}$) v. Disallowed-Realized Ratio. RIGHT: Average Size of Original Agent Neighborhoods ($S\textsubscript{O}$) v. Disallowed-Realized Ratio. The red plots in both panels represent the Gradual Migration scenario where $r\textsubscript{mig} = 0.005\%$ over all 500 iterations, while the blue plots represent the Rapid Migration scenario, with $r\textsubscript{mig} = 0.025\%$ over 100 iterations and $r\textsubscript{mig} = 0.0\%$ over the remaining 400 iterations.}
\label{fig1}
\end{figure}

We begin with the outcomes on the `dual' segregations. When we study the behavior of Average Fraction of Rich Neighbors ($F\textsubscript{R}$) with decreasing $\beta\textsubscript{move}$, we expect to see a decreasing trend as movement becomes easier in contravention of the wealth threshold condition, therefore making it progressively more difficult for richer agents to congregate spatially. Additionally, we also expect to see a sharp transformation from a segregated to mixed wealth state, as observed in our previous work~\cite{bib0b2}. The left panel of Figure~\ref{fig1} plots $F\textsubscript{R}$ against the Disallowed-Realized Ratio, which is the fraction of disallowed-realized moves to total agent moves, and indeed finds this sharp transformation under both migration scenarios. It is, however, important to note that the onset of the sharp transformation is happening much later now, at a Disallowed-Realized Ratio of $\sim$36\%, when compared to our earlier model~\cite{bib0b2}, where it happened when the Disallowed-Realized Ratio increased just above 0. 

Average Size of Original Agent Neighborhoods ($S\textsubscript{O}$) also shows a non-linear sharp transformation, but from a less identity-segregated state to a more identity-segregated state with declining $\beta\textsubscript{move}$. Therefore, as $\beta\textsubscript{move}$ decreases, increasing the ease of movement by progressively relaxing the wealth threshold condition and thus causing a decline in wealth based segregation, it simultaneously appears to exacerbate identity based segregation. Observing the right panel of Figure~\ref{fig1}, we notice that the onset of the  sharp transformation in $S\textsubscript{O}$ occurs at a Disallowed-Realized Ratio of $\sim$36\%, and peaks at a Disallowed-Realized Ratio of $\sim$45\% in all scenarios. We discuss the potential rationale for these opposing segregations in the next section.

\begin{figure}[ht]
\centering
\includegraphics[width=\linewidth]{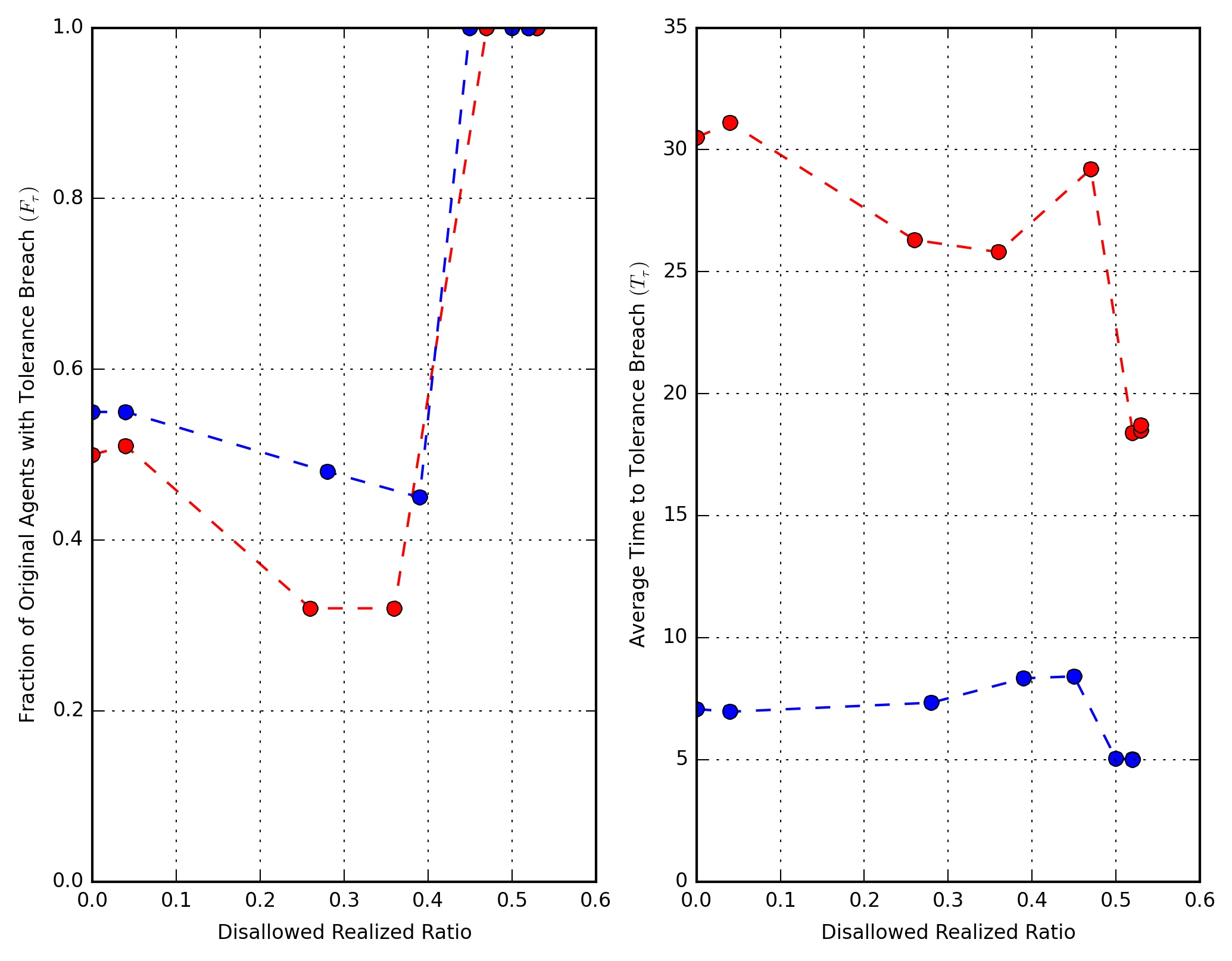}
\caption{Impact of migration rate on migrant tolerance levels ($\tau$) of original agents. LEFT: Fraction of Original Agents with Tolerance Breach ($F\textsubscript{$\tau$}$) v. Disallowed-Realized Ratio. RIGHT: Average Time to Tolerance Breach ($T\textsubscript{$\tau$}$) v. Disallowed-Realized Ratio. The red plots in both panels represent the Gradual Migration scenario where the migration rate $r\textsubscript{mig} = 0.005\%$ over 500 out of 500 iterations, while the blue plots represent the Rapid Migration scenario, with migration rate $r\textsubscript{mig} = 0.025\%$ over 100 out of 500 iterations and $r\textsubscript{mig} = 0.0\%$ over the remaining 400 iterations.}
\label{fig2}
\end{figure}

We now turn to the measures relating to the impact of migration rates on migrant tolerance levels of original agents. For a Disallowed Realized Ratio $\geq$$\sim$45\%, the average Fraction of Original Agents with Tolerance Breach ($F\textsubscript{$\tau$}$) is $\sim$100\%, while below this ratio, $F\textsubscript{$\tau$}$ is in the region 35\%-50\%. The left panel of Figure~\ref{fig2} confirms this non-linear increase in  $F\textsubscript{$\tau$}$ as the ease of agent movement increases with decreasing $\beta\textsubscript{move}$.

As demonstrated in the right panel of Figure~\ref{fig2}, the difference in the Average Time to Tolerance Breach ($T\textsubscript{$\tau$}$) between the  gradual and rapid migration scenarios is quite substantial. Original agent tolerance is breached significantly faster under rapid migration. For instance, when $\beta\textsubscript{move}\geq10$, we observe that $T\textsubscript{$\tau$}$ for the rapid migration scenario is $\sim7$ time sweeps as against $\sim$29 time sweeps under the gradual migration scenario, a 4.1 times faster attainment of tolerance breach.

\section*{Discussion}
We structure this discussion around three significant themes that emerge from the analysis: (i) the point of onset of dramatic transformation in wealth based segregation; (ii) the emergence of simultaneous non-linear transformations in wealth- and identity-based segregation; and (iii) the difference in velocity to tolerance breach under rapid and gradual migration.

We have previously explained~\cite{bib0b2} the sharp transformation from segregated to mixed wealth neighborhoods on account of the entropic effect created by even a few disallowed-realized moves, which enable the creation of neighborhood wealth configurations that cause a non-linear increase in moves satisfying the wealth threshold condition (allowed moves), such that the overall effect is the emergence of a rapid de-segregation. As we see in the left panel of Figure~\ref{fig1}, the onset of rapid de-segregation is at a Disallowed-Realized Ratio of $\sim36$\%, compared to our earlier model, where the onset was at just above 0\%. We find that this behavior is explained by the extent to which agents are willing to attempt a move to neighborhoods with lower median wealths, an aspect mediated by $\beta\textsubscript{choice}$ in the calculation of $p\textsubscript{choice}$. The more agents that are willing to move in contravention of this neighborhood comparison condition, the earlier is the onset of de-segregation. This argument is substantiated by our analysis of outcomes at different $\beta\textsubscript{choice}$ values (detailed in Supplementary Information Appendix (SI.4)): we find the onset of the sharp transformation at a lower value of Disallowed-Realized Ratio ($\sim20$\%) for $\beta\textsubscript{choice} = 2$, and at a higher value of Disallowed-Realized Ratio ($\sim41$\%) for $\beta\textsubscript{choice} = \infty$. Therefore, while the sharpness of transformation from segregated to mixed-wealth states is evident in all scenarios, the actual onset of the transformation requires progressively greater disallowed-realized moves as the propensity of agents to move out of their extant neighborhoods in contravention of the neighborhood comparison condition decreases. Figure~\ref{fig3} presents the curves for different $\beta\textsubscript{choice}$, illustrating our argument:

\begin{figure}[ht]
\centering
\includegraphics[width=\linewidth]{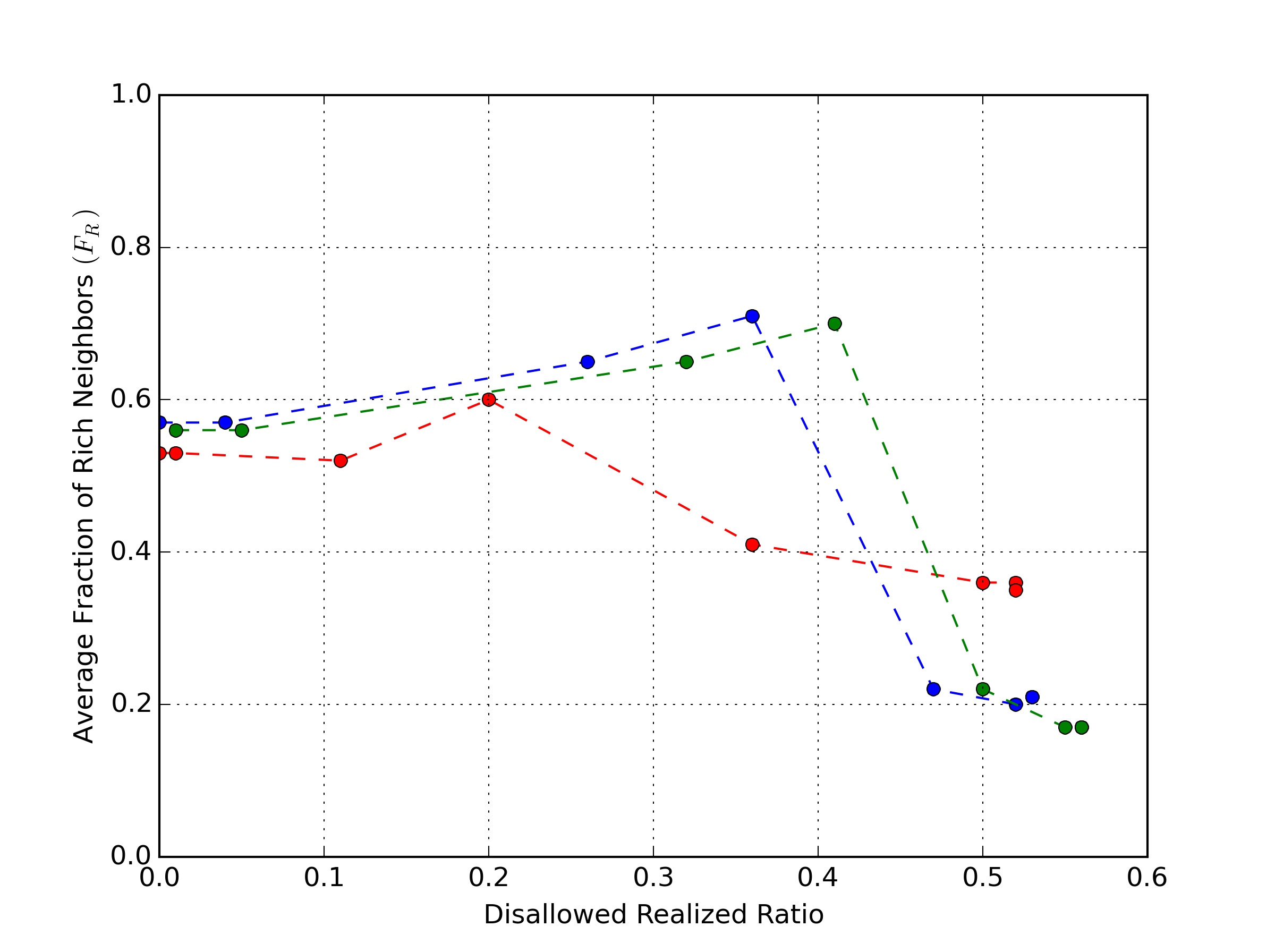}
\caption{Average Fraction of Rich Neighbors ($F\textsubscript{R}$) v. Disallowed-Realized Ratio. The red curve describes model behavior at $\beta\textsubscript{choice}$ = 2, the blue curve at $\beta\textsubscript{choice}$ = 10, and the green curve at $\beta\textsubscript{choice}$ = $\infty$. }
\label{fig3}
\end{figure}

We now explore the emergence of the dual sharp transformations in spatial configurations of agent wealth and of identity. Just as there is the onset of transformation from a segregated to a mixed wealth state, we see a corresponding sharp transformation from a reasonably mixed identity configuration to an increasingly segregated identity configuration. It is important to note that even with the reasonably mixed identity configuration at high $\beta\textsubscript{move}$, when agent movement is restricted, some amount of identity based clustering has nevertheless been established. A completely mixed identity-based configuration, for instance, would yield an Average Size of Original Agent Neighborhoods, $S\textsubscript{O}(mix)$ = 49, which is indeed, on average, the value of $S\textsubscript{O}$ at the start of our simulations. For  $\beta\textsubscript{move}\geq100$, we see from the right panel of Figure~\ref{fig1} that $S\textsubscript{O}$ is $\sim212$, which is $\sim4.3$ times $S\textsubscript{O}(mix)$, reflecting the fact that there is a not insignificant amount of identity-based clustering that has occurred even at this stage. We also observe from the left panel of Figure~\ref{fig2} that the average Fraction of Original Agents with Tolerance Breach ($F\textsubscript{$\tau$}$) is $\sim51\%$, for $\beta\textsubscript{move}\geq100$. This essentially means that for $\beta\textsubscript{move}\geq100$, $\sim49$\% of the original agents have not seen their migrant tolerance condition breached, implying that this fraction of agents have been able to spatially organize themselves in a manner that keeps migrants away or slows their entry into these neighborhoods. Given that original agents are drawn from a higher wealth distribution, their wealths are, on average,  greater than those of migrants. We would therefore expect to see early on in the dynamics that the richer original agents, especially, are able to segregate themselves and create rich neighborhoods, thus erecting substantial wealth thresholds preventing significant entry by migrants unless they are able to afford it. But given the discrepancies in wealth of original agents and migrants, we would expect that the numbers of migrants entering these rich cells is very low and even over time their growth is not able to overshadow the population of original agents. We get confirmation of this when we analyze the fraction of original agents in neighborhoods and find that original agents in the richest 20\% of neighborhoods comprise $\sim75\%$ of the total population of original agents, and also that this subset of original agents represents $\sim83\%$ of those who have not seen their tolerance breached. Additionally, the fact that wealth based segregation is significant for $\beta\textsubscript{move}\geq100$ implies that the original agents in these neighborhoods are, on average, rich agents. Migrants are more easily able to enter the rest of the neighborhoods, and the entry of some migrants early on makes it possible over time for greater numbers of migrants to enter these neighborhoods even if in contravention of the wealth condition. We would therefore expect that a significant proportion of the migrant agents are in the poorer neighborhoods. This is confirmed when we analyze the fraction of migrant population in the rich and poor neighborhoods: migrant agents in the richest 20\% of neighborhoods comprise $\sim56\%$ of the total population in these neighborhoods, which represents $\sim20\%$ of the overall migrant population; however, in the poorest half of neighborhoods, they comprise $\sim99\%$ of the total population, with original agents almost completely absent from these cells.

As disallowed-realized moves increase (with decreasing $\beta\textsubscript{move}$), and at a Disallowed-Realized Moves ratio of $\sim36\%$, we see the sharp rise in identity based clustering, coinciding with the onset of dramatic drop in wealth based segregation, as evidenced by the two panels of Figure~\ref{fig1}. The emergence of more mixed wealth configurations increasingly allows lower wealth migrants to move into neighborhoods they were earlier unable to afford. Specifically, the dynamics allow more of the poorer original agents to cluster with the richer original agents in contravention of the neighborhood wealth threshold condition, thus not only creating mixed wealth neighborhoods but also increasingly large clusters of original agents. This argument is substantiated by the population profile of the richest 20\% of neighborhoods (for $\beta\textsubscript{move}\leq 5$), where we find over 99\% of all original agents situated - thus generating the high levels of identity based segregation. However, in contrast to the case when $\beta\textsubscript{move}\geq 100$, where only $\sim20\%$ of the migrant population was situated in the richest 20\% of neighborhoods, for $\beta\textsubscript{move}\leq 5$ we find that this has gone up to  $\sim69\%$ - thus generating the mixed wealth configurations yielding the essential decline in wealth-based segregation. 

Therefore, our analysis suggests that given a society where the two forces of wealth- and identity-based spatial clustering are in effect, we would indeed expect them to work in opposite directions - when wealth-based segregation breaks down and weakens, identity-based segregation emerges and strengthens, and conversely, when wealth segregated configurations emerge, we see much lower levels of identity-based segregation (though still at levels significantly above completely mixed identity configurations). In Schelling's original model~\cite{bib1}, we see the emergence of very high levels of segregation (81.5\%) at agent tolerance level = 0.5, in the absence of a wealth threshold. This leaves us with the conclusion that the restrictions on movement imposed by neighborhood wealth thresholds (at high $\beta\textsubscript{move}$) indeed controls the extent of identity-based segregation, which in the absence of these restrictions could be substantially higher. Additionally, we can also infer that when agents have even a slight preference for other agents with like identity, it is practically impossible to get completely mixed identity configurations and that some amount of identity based segregation is inevitable, irrespective of the stringency of the wealth threshold condition.

Our analysis also indicates that the mixed wealth configurations created by increasing disallowed-realized moves directly enable the corresponding sharp transformation from a less identity-segregated to highly identity-segregated city. We can think of the highly mixed wealth configurations as scenarios where neighborhood wealth thresholds essentially become superfluous ($\beta\textsubscript{move}\leq 0.1$) and it is in these cases we see the full extent of possible identity-based segregation because original agents congregate in neighborhoods they prefer with practically no wealth restrictions on them. It is therefore the case the very movement of agents that allow for decreased wealth segregation, ends up enabling the exacerbation of identity-based segregation.

The reversal of segregations based on wealth and on identity (such as race/caste/ethnicity/religion) in conditions of increasing migration are important public policy considerations in many urban societies around the world. Our work suggests that improving segregation outcomes on one of these fronts can end up exacerbating outcomes on the other. This understanding could be of use to policy makers in assessing the potential impact of any socio-economic policy on the fundamental trade-off between these dual segregation tendencies.

Finally, we turn to the implications of Average Time to Tolerance Breach ($T\textsubscript{$\tau$}$) under the gradual and rapid migration scenarios. Realistically, we can think of rapid migration scenarios as representative of sudden increases in population over short time periods as caused, for example, by migrants and refugees fleeing conflict or natural catastrophe. As illustrated in Figure~\ref{fig1}, over long time periods, there appears to be no difference in the patterns of segregation under the gradual and rapid migration scenarios. This is in agreement with Urselmans' findings~\cite{bib0b4} that the rate and quantum of immigration do not impact outcomes in the long term. Our interest though, is also on the short term impacts, and this is demonstrated in the right panel of Figure~\ref{fig2} - $T\textsubscript{$\tau$}$ in the case of rapid migration is significantly lower than in the gradual migration scenario.  This sharp difference occurs even as the left panel of Figure~\ref{fig2} demonstrates the relative similarity in behavior of $F\textsubscript{$\tau$}$ under both rapid and gradual migration scenarios. This is because rich residents are able to segregate themselves quickly (as we discussed earlier), erecting neighborhood wealth barriers that prevent migrants from entering in large numbers. However, the remarkable difference in $T\textsubscript{$\tau$}$ is on account of the lack of time available, in the face of the sharp, significant burst in migration, for other residents to spatially realign themselves in configurations that enable them to meaningfully delay the time to tolerance breach, resulting in a significantly quicker breach of $\tau$. For $\beta\textsubscript{move}\ge 10$, which we regard as reasonable representations of wealth thresholds in real world cities~\cite{bib0b3}, $T\textsubscript{$\tau$}$ under gradual migration is over 4 times that under rapid migration. Therefore, while long-term outcomes under these migration scenarios are very similar, there are significant differences in near-term outcomes. The fact that agents have their tolerance limits breached in short time spans could potentially underpin the intensity of response of original agents to this abrupt change in the composition of their neighborhoods. This is to say that the response of residents to sudden, sharp spikes of migration which result in a significantly faster breach of $\tau$ could therefore be much more urgent and of higher intensity than if the same population of migrants entered over a longer period of time (yielding a much higher $T\textsubscript{$\tau$}$). It is a possibility that this phenomenon of rapid tolerance breach is what underlies the nature of resident responses to sharp migrations, as described for example by Wike, Stokes and Simmons~\cite{bib0ja} in their survey of European attitudes to Syrian refugees.

\section*{Conclusion}
We build on our previous work on wealth segregation by incorporating migration and explore the emergence of and interplay between identity-based and wealth-based segregations. We find that the sharp transformation we observed between segregated and mixed-wealth states earlier obtains here as well, but that it is accompanied by a corresponding sharp transformation from a less identity-segregated to a highly identity-segregated state. We argue that it is the mixed wealth configurations that are enabled by an easing of the neighborhood wealth threshold condition, which provide the impetus for increased identity-based segregation. Therefore, our work suggests that a decrease in wealth segregation does not merely accompany, but in fact drives, the increase in identity based segregation. From a policy perspective, it could be important to recognize this fundamental trade-off in designing socio-economic programs aimed at addressing the problem of segregation. Additionally, we also find that the time taken for migrant tolerance levels to be breached for a plurality of agents is substantially lower in case of rapid migrations over short time spans than when the migration is more gradual, even though population levels and segregation outcomes under these scenarios are similar over longer time spans. This speed to tolerance level breach could underlie the intensity of resident responses to sharp bursts of migration.

\bibliography{sample}

\section*{Acknowledgments}

AS gratefully acknowledges the financial support received in the form of the Schrödinger Scholarship from the Faculty of Natural Sciences, Imperial College London. The funders had no role in this research study.

\section*{Author Contributions Statement}

A.S. and H.J.J. conceptualized the research, chose the methodology, performed the formal analysis, and reviewed and edited the manuscript. A.S. programmed the simulation and wrote the draft manuscript.

\section*{Additional information}

\textbf{Competing financial interests}:
The authors declare no competing financial interests.
\newline
\noindent
\textbf{Data availability statement}:
All data was simulated based on the model description in `Model definition and specifications' section. 

\end{document}